\documentstyle[epsfig,aas2pp4]{article}

\begin{document}

\title{Changes in the long term intensity variations in Cyg~X-2~and~LMC~X-3}

\author{B. Paul$^{1,}\footnotemark[3]$, S. Kitamoto$^2$ and F. Makino$^1$ }
\affil{$^1$Institute of Space and Astronautical Science,
3-1-1 Yoshinodai, Sagamihara, Kanagawa 229-8510, Japan}
\affil{$^2$Department of Earth and Space Science, Faculty of Science,
Osaka University 1-1, Machikaneyama, Toyonaka, Osaka, 560, Japan}
\affil{e-mail:~~bpaul@astro.isas.ac.jp, kitamoto@ess.sci.osaka-u.ac.jp,
makino@astro.isas.ac.jp}
\footnotetext[3]{On leave from the Tata Institute of Fundamental
Research, Homi Bhaba Road, Mumbai, 400005, India}

\begin{abstract}

We report the detection of changes in the long term intensity
variations in two X-ray binaries Cyg X-2 and LMC X-3. In this work, we have
used the long term light curves obtained with the All Sky Monitors (ASM) of
the Rossi {\it X-ray Timing Explorer (RXTE), GINGA, ARIEL 5} and {\it VELA 5B}
and scanning modulation collimator of {\it HEAO~1}. It is found that in the
light curves of both the sources, obtained with these instruments at various
times over the last 30 years, more than one periodic or quasi-periodic
components are always present. The multiple prominent peaks in the
periodograms have frequencies unrelated to each other. In Cyg X-2, {\it
RXTE-ASM} data show
strong peaks at 40.4 and 68.8 days, {\it GINGA-ASM} data show strong peaks at
53.7 and 61.3 days. Multiple peaks are also observed in LMC X-3. The various
strong peaks in the periodograms of LMC X-3 appear at 104, 169 and 216
days with {\it RXTE-ASM}, and 105, 214 and 328 days with {\it GINGA-ASM}. The
present results, when compared with the earlier observations of periodicities
in these two systems, demonstrate the absence of any stable long period. The
78 day periodicity detected earlier in Cyg X-2 was probably due to the
short time base in the {\it RXTE} data that were used and the periodicity of
198 days in LMC X-3 was due to a relatively short duration of observation with
{\it HEAO~1}.

\end{abstract}


\section{Introduction}

Many X-ray binaries are highly variable
in their X-ray intensity over long time scales. But in most of the sources,
either the intensity variations are aperiodic or their periodic nature has not
yet been discovered because of long-period, low modulation or a lack of
sensitive uninterrupted monitoring. The All Sky Monitor (ASM) onboard the Rossi
{\it X-ray Timing Explorer (RXTE)} has produced light curves of many bright
X-ray binaries for about 1,000 days with large signal to noise ratio. The
{\it RXTE-ASM} data along with the {\it GINGA-ASM, VELA 5B, ARIEL 5} and
{\it HEAO~1} data have a large time base and it is possible not only to
search for any periodic or quasi periodic intensity variations of a few days
to a few months time scale, but also to investigate changes in
the timing behaviour. Changes in the long term periodicity have been observed
in SMC X-1 (Wojdowski et al. 1998) and GX 354-0 (Kong et al. 1998). We
selected Cyg X-2 and LMC X-3 to study any possible changes in the long
term periodicity because in these two sources long term periodicity is
known to exist in data bases covering a very long time and also for the
fact that light curves for the intervening period 1987--1991 were available
from the {\it GINGA-ASM}.

Cyg X-2, a bright X-ray source, was discovered by Byram et al. (1966)
with a sounding rocket experiment. A binary period of 9.8 days, other
orbital parameters and mass limits of the two components of this binary
system were measured by Cowley et al. (1979). This is a low mass X-ray
binary and a typical Z source that shows Type-I X-ray bursts (Kahn \&
Grindlay 1984; Smale 1998), 18$-$50 Hz QPOs in the horizontal branch
and 5.6 Hz QPOs in the nominal branch (Hasinger et al. 1986; Elsner et
al. 1986; Wijnands et al. 1997; Kuulkers, Wijnands and van der Klis 1999).
Recently, with the {\it RXTE}, kHz
QPOs with two peaks have also been detected from this source (Wijnands
et al. 1998). Vrtilek et al. (1988) discovered X-ray dips of different
types and some were found to occur in one particular phase of the binary
period. A 77 day periodicity was discovered from the {\it VELA 5B }
observations (Smale \& Lochner 1992). Wijnands, Kuulkers \& Smale (1996)
reported detection of a periodicity of 78 days from the first 160 days
of ASM data, and showed that their result was supported by the archival
data of {\it ARIEL 5} and {\it VELA 5B}.

LMC X-3, a high mass X-ray binary, was discovered with the {\it UHURU}
satellite (Leong et al. 1971) and its position was measured accurately with
the {\it HEAO~1} scanning modulation collimator (Johnston et al. 1978). From
spectroscopic observations Cowley et al. (1983) discovered an orbital
period of 1.7 days and a mass function of 2.3 M$_\odot$. LMC X-3 is
considered to be a very strong black hole candidate (BHC) due to the fact that
the mass of the compact object has a lower limit of 9 M$_\odot$. But, {\it
HEAO~A2} observations revealed that unlike other known black hole candidates,
LMC X-3 lacks rapid X-ray variability (Weisskopf et al. 1983). The lowest
time scale for 1\% rms amplitude variation was derived from the {\it EXOSAT}
observations to be 600~s (Treves et al. 1988). One possible
explanation for the lack of rapid X-ray variability is that most of the
time LMC X-3 is found to be in a high state with an unusually soft X-ray
spectrum (White \& Marshal 1984), and the rapid X-ray variations are
generally subdued in Cyg X-1 like BHCs in their high state. {\it GINGA}
observations showed that the energy spectrum consists of a soft, thermal
component and a hard, power-law component. In spite of large changes in
mass accretion rate and disk temperature, the inner radius of the
accretion disk was found to be remarkably constant and was suggested to
be related to the mass of the compact object (Ebisawa et al. 1993).
Long term intensity variation is well known in LMC X-3 and Cowley et
al. (1991) discovered a periodicity of 198 (or 99) days from
the {\it HEAO~A1} and {\it GINGA} large area counter data.

The light curves of Cyg X-2 and LMC X-3 obtained with the ASMs onboard the
{\it RXTE, GINGA, VELA 5B} (only for Cyg X-2), {\it ARIEL 5} and the scanning
modulation collimator of the {\it HEAO~1} (only for LMC X-3) have been used
to investigate the long term intensity variations in these two sources and
we have found presence of multiple components of flux variations in both the
sources. The data used here together covers a time base of 30 years for
Cyg X-2 and 26 years for LMC X-3.

\section{Data and analysis}

The ASM on board {\it RXTE} scans the sky in a series of dwells of about 90 s
each, 
and any given X-ray source is observed in about 5$-$10 such dwells every day.
The details about the ASM detectors and observations with the ASM are
given in Levine et al. (1996). We have used the quick look ASM data
obtained in the period 1996 February 20 to 1999 February 4, provided by the
{\it RXTE-ASM} team. The ASM data are available in two different forms, per
dwell and one day average. The periodograms obtained from the
two sets of data are identical except for a normalization. Hence the
results obtained from the one day average data are presented here. The {\it
GINGA-ASM} was operational during 1987 to 1991 and details of the detectors,
operations, and the detection techniques have been described by Tsunemi et
al. (1989). The all sky observations were performed about once every day when
the satellite was given one full rotation with about 70\% sky coverage by
the ASM. The detection limit in one such rotation was of the order of 50
mCrab and was dependent on the position of the source with respect to the
satellite equator. Archival data of {\it VELA 5B} and {\it ARIEL 5} were
obtained from the HEASARC data base and {\it HEAO~1} data for LMC X-3 were
taken from previously published work (Cowley et al. 1991). For details about
the {\it ARIEL 5} ASM and the {\it VELA 5} instrument please refer to
Holt (1976) and Priedhorsky, Terrel \& Holt (1983) respectively.

To search for periodicities in the unevenly sampled data, we have used the
method suggested by Lomb (1976) \& Scargle (1982). For Cyg X-2, we generated
periodograms in the period range of 10$-$160 days from the {\it RXTE-ASM,
GINGA-ASM, ARIEL 5} and {\it VELA 5B} data. The {\it ARIEL 5, VELA 5B} and
initial part of the {\it RXTE-ASM} light curves were analyzed earlier and
long periods of 78 and 69 days were reported (Smale \& Lochner 1992;
Wijnands et al. 1996, Kong et al. 1998). For LMC X-3, light curves from
{\it RXTE-ASM, GINGA-ASM, ARIEL 5} and {\it HEAO~1} were used to generate
periodogram in the range of 20$-$500 days. {\it VELA 5B} data for LMC X-3
has low signal to noise ratio and is not used here. The discovery of a 198
(or 99) day periodicity from the {\it HEAO 1} light curve was made by Cowley
et al. (1991), and the initial part of the {\it RXTE-ASM} data also showed
similar variation (Wilms et al. 1998). The period ranges chosen for the
periodogram analysis are such that a sub-harmonic or first harmonic of the
earlier known periods can be identified for the respective sources. For
Cyg X-2, we also verified that in the 160-500 days range the periodograms
generated from data sets are featureless except for {\it ARIEL 5} which
shows peaks around 180 and 360 days due to known yearly effect (Priedhorsky
et al. 1983).

\begin{figure}[t]
\centerline{\psfig{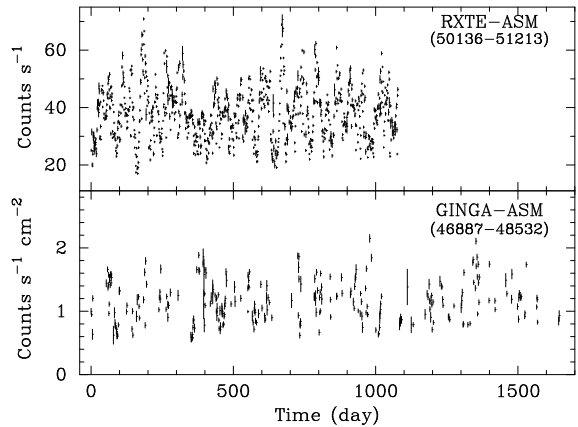}}
\caption{The {\it RXTE-ASM} (1.5--12 keV) and {\it GINGA-ASM}
(1--20 keV) light curves of Cyg X-2. The days of the observations are given
in truncated Julian days in the figure.}
\end{figure}

The significance of the various peaks detected in the periodograms is given
in Table 1 in terms of the false alarm probability (FAP) calculated following
the
method suggested by Horne \& Baliunas (1986). The window functions, which
can induce artificial periodicities were calculated for the different time
series used here. The {\it GINGA-ASM} time series, which is very susceptible
to artificial periodicities because of its scarce sampling, does not show
any alarming feature in the period
range used in this paper and the effect of the window function is less for
the other instruments. However, we wish to point out that the use of
window function to find spectral leakage can sometimes be misleading for this
kind of light curves. As an example, in the {\it HEAO 1} light curve of LMC X-3,
the density of data points available is larger when the source is brighter,
and this is likely to be true for other sources and instruments also. This
results in some features in the window function at periods near 80 and 160 days,
whereas a glance at the light curve (Figure 4) leaves no doubt about the
presence of a periodic variation. We have verified the significance of the
peaks in the periodograms independently using two more methods. Following
Kong et al. (1998), we have generated light curves using random numbers, with
the time series, average, and variance similar to the real light curves and
calculated the periodograms for 10,000 such light curves. The highest points
in these periodograms were identified and a power which is larger than the
highest points of 99\% of the periodograms is considered to correspond to a
false alarm probability of 10$^{-2}$. The same process was carried out also
for 10,000 light curves with the same time series as the real light curves,
but the count rates redistributed randomly. The results from these two
analysis are found to be identical and the 10$^{-2}$ false alarm levels are
indicated by dashed lines in all the periodograms. Absence
of strong peaks in the periodograms calculated from the simulated and
redistributed light curves confirms that the peaks observed in the periodograms
of the real light curves are not artifact of observation windows.

\subsection{Cyg X-2}

Presence of strong intensity variations by a factor of $\sim$2--4 on time
scale of weeks is well known in this source (Kuulkers, van der Klis, \&
Vaughan 1996; Wijnands et al. 1996) and can be clearly seen in the light
curves of {\it RXTE-ASM} and {\it GINGA-ASM} (Figure 1). A periodic nature of
this intensity variation with period of less than 100 days is also apparent
in the light curves.

The periodograms obtained from the {\it RXTE-ASM, GINGA-ASM, ARIEL 5} and
{\it VELA 5B} data are shown in Figure 2. In the RXTE-ASM data, there are
several significant peaks and the two most prominent ones are at 40.4 and
68.8 days indicating the presence of multiple periodicities in this system.
In the {\it GINGA-ASM} periodogram there are two prominent peaks at
53.7 and 61.3 days and there are several smaller peaks. The periodogram of
the {\it ARIEL 5} data is a very complicated one and has many prominent
peaks at frequencies ranging between 30 and 80 days. In addition to
the 78 days peak noticed by Wijnands et al. (1996), there are other peaks at
41, 46 and 54 days.  The {\it VELA 5B} data has low signal to noise
ratio and the power in the periodogram is much smaller, but two peaks at 33.9
and 77.4 days can be clearly seen in Figure 2. The significance of the peaks
detected in various periodograms are given in Table 1.

\begin{figure}[t]
\centerline{\psfig{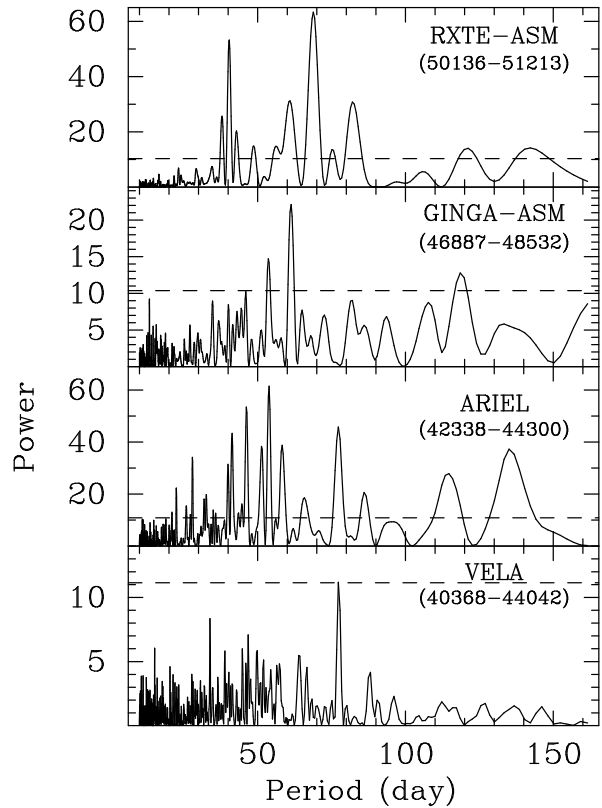}}
\caption[fig2.ps]{The Lomb-Scargle periodograms obtained from the
{\it RXTE-ASM, GINGA-ASM, ARIEL 5} and {\it VELA 5B} light curves of
Cyg X-2. The horizontal dashed line represent the 99\% confidence limits
(see text for details).}
\end{figure}

To investigate whether these multiple peaks in the periodograms originate
in different parts of the light curves or whether multiple peaks are present
throughout the entire light curve of each satellite, we have divided the
light curves
of {\it RXTE} and {\it ARIEL 5} into three equal segments and generated
periodograms from each of them.  There are several prominent peaks in each of
the periodograms, and the periods of these peaks are noted in Table 1 along
with the significance of their detection. Periodograms obtained from the three
358 day segments of the {\it RXTE} light curve are shown in Figure 3
along with the respective light curves. It is evident from Figure 3 and Table
1 that there are multiple periodicities in this system that are uncorrelated
and have varying amplitude and period. The higher frequency variation at
around 40 days time scale appears to be more stable in frequency than the
lower frequency variations around 70 days during these observations.

\begin{figure}[t]
\centerline{\psfig{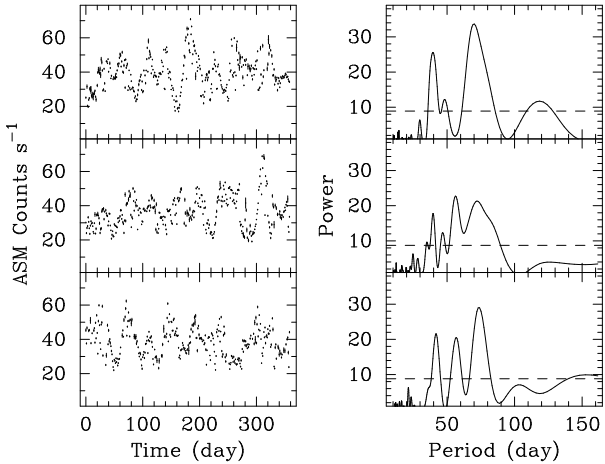}}
\caption[fig3.ps]{Three segments of the {\it RXTE-ASM} light curve of
length 358 days each and the corresponding Lomb-Scargle periodograms are
shown on the left and right hand sides respectively. The 99\% confidence
limits are shown with the dashed horizontal lines.}
\end{figure}

\subsection{LMC X-3}

Long term light curves of LMC X-3, obtained with the {\it RXTE-ASM, GINGA-ASM}
and {HEAO~1} are shown in Figure 4. The {HEAO~1} data, plotted in the lower
panel of the figure, very clearly shows almost periodic intensity variations
at about 100 days. However, as Cowley et al. (1991) pointed out, the intensity
modulation is missing during the first 100 days. They concluded from this data
that the intensity variations in LMC X-3 is periodic at $\sim$ 198 (or
possibly $\sim$ 99) days. The {\it GINGA-ASM} light curve shown in the middle
panel of the figure also indicates intensity modulations, but at a larger time
scale of about 200 days. Two such modulations can be clearly identified at the
end of the light curve and one more in the middle. There are also some episodes
of about 100 days periodic variations in some parts of the {\it GINGA-ASM} light
curve. The {\it RXTE-ASM} light curve of LMC X-3, as shown in the top panel
of Figure 4, shows a varying nature of the approximately 100 or 200 days
intensity variations. At the beginning, there are three strong modulations at
about 100 days, similar to what was seen with {\it HEAO~1}, but in the later
part of the light curve, the modulations are much less prominent and have
larger time scale. Intensity variations in LMC X-3 are not clearly visible in
the {\it ARIEL 5} light curve (not shown here) because of sparse sampling.

\begin{figure}[t]
\centerline{\psfig{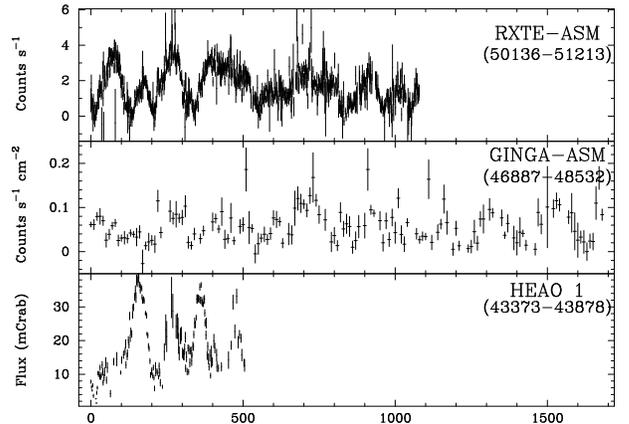}}
\caption[fig4.ps]{The long term light curves of LMC X-3 obtained with
the {\it RXTE-ASM, GINGA-ASM} and {\it HEAO~1} satellites. The energy ranges
are 1.5--12 keV, 1--20 keV and 1--13 keV respectively.}
\end{figure}

In Figure 5 we have shown the periodograms generated from the light curves of
LMC X-3 obtained with all the four satellites mentioned above. There are three
prominent peaks at 104.4, 168.8 and 215.6 days in the periodogram obtained
from the {\it RXTE-ASM} light curve. The peak at 104.4 days is narrow while
the other two are broad. In the periodogram of the {\it GINGA-ASM}, however,
there are two prominent peaks at 214 and 328 days. There is also a less
significant indication of some periodic component at 105 days. The {\it HEAO~1}
data which have been extensively discussed by Cowley et al. (1991), predictably
shows two peaks at 99 and 203 days with the former being narrower. Results
from the {\it RXTE} and {\it HEAO~1} are similar in nature except for the fact
that the broad peak near 200 days is resolved into two components in the {\it
RXTE-ASM} data. This also shows that the two peaks near 100 and 200 days as
seen in the periodogram of the {\it HEAO~1} data are not related. The
periodogram of the {\it GINGA-ASM} light curve also has a shape similar to the
other two but the time scale is a factor of two larger. The {\it ARIEL 5}
light curve has low signal to noise ratio and infrequent sampling and the
periodogram obtained from this data shows three less significant peaks at 90,
96 and 130 days. At periods below 50 days, this periodogram is very noisy.
From the four periodograms shown in Figure 5 it appears that there is a quasi
periodic component at around 100 days in LMC X-3 whose strength is time
dependent. There is at least one more component at longer period of about
200 days. The period excursion of the $\sim$100 days component is significant,
but relatively smaller than the other periodicities.

\begin{figure}[t]
\centerline{\psfig{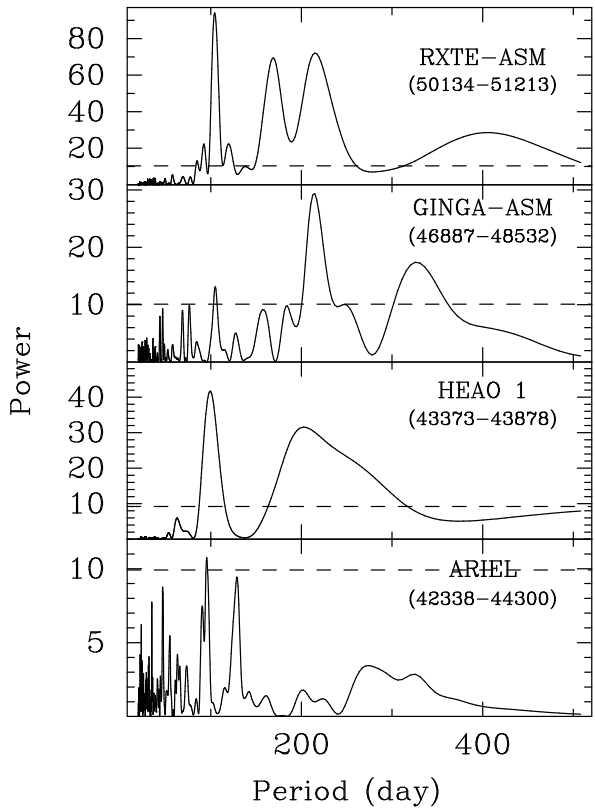}}
\caption[fig5.ps]{The Lomb-Scargle periodograms generated from the light
curves of LMC X-3 obtained with the {\it RXTE-ASM, GINGA-ASM, HEAO~1} and
{\it ARIEL 5} detectors. The dashed horizontal lines indicate the 99\%
confidence limits.}
\end{figure}

To investigate the nature of the multiple components of intensity variations
in more detail, we have done further analysis of the {\it RXTE-ASM} data of
LMC X-3 in a manner similar to what was done with the Cyg X-2 data. We divided
the light curve into two segments, each 540 days long, and have generated the
periodograms from both of the segments that are shown in Figure 6.
Two large peaks at 104.3 and 177.8 days are clearly seen in the first
periodogram whereas the periodogram generated from the second part of the
light curve shows two peaks at 101.9 and 221.6 days with much less strength.
It appears that the $\sim$104 day periodicity is suppressed in the second
part of the {\it RXTE} data and the second component of intensity variation
has moved to a longer time scale. The significance of the peaks in the
periodograms are given in Table 1.

\begin{figure}[t]
\centerline{\psfig{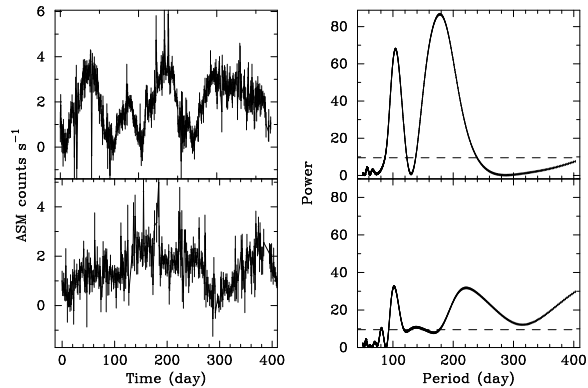}}
\caption[fig6.ps]{The Lomb-Scargle periodograms obtained from two segments
of the {\it RXTE-ASM} light curve of LMC X-3. Each segment of the light curve
is of 540 days duration. The dashed horizontal lines indicate the 99\%
confidence limits.}
\end{figure}

\subsection{Spectral variations}

In Cyg X-2, the hardness ratio and intensity that define its position in
the Z track and also the position of the Z track itself, changes significantly
at short time scales. But at longer time scales, in between different
observations, a negative correlation was found between the hardness ratio and
total intensity with {\it EXOSAT} (Kuulkers et al. 1996), {\it GINGA}
(Wijnands et al. 1997) and also with the {\it RXTE-ASM} during its first few
months observations (Wijnands et al. 1996). The hardness ratio in LMC X-3, on
the other hand was found to have positive correlation with intensity (Cowley
et al. 1991). We have calculated the two hardness ratios HR1 (3.0-5.0
keV/1.5-3.0 keV) and HR2 (5.0-12 keV/3.0-5.0 keV) from the {\it RXTE-ASM}
data as a function of the total intensity for the two sources (Figure 7)
using all the available data. In Cyg X-2, HR2 is negatively correlated
with luminosity (correlation coefficient -0.6 and probability of no correlation
10$^{-30}$) but HR1 does not show any correlation (coefficient -0.04,
probability 0.5). If the data points are connected by lines, the HR2 plot
appears like a loop indicating that HR2 follows two different tracks during
the rising and decaying phases of the intensity variations. Large deviation
in the hardness plot of Cyg X-2 is due to movement of the source along the Z
track (Wijnands et al. 1996) and changing position of the Z track in the
color-color diagram. In LMC X-3, a weak positive correlation is
found for HR1 (0.5, 10$^{-5}$) but no correlation for HR2 (-0.07, 0.5). The
correlation coefficients and probabilities were calculated using two different
methods: the linear and rank correlation, both gave identical results.

\begin{figure}[t]
\centerline{\psfig{figure=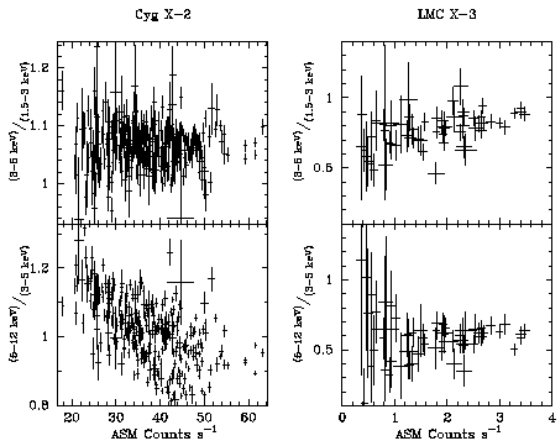,width=3.2in,angle=00}}
\caption[fig7.ps]{Hardness ratios HR1 (3-5 keV / 1.5-5 keV) and HR2
(5-12 keV/ 3-5 keV) of Cyg X-2 and LMC X-3 obtained from the {\it RXTE-ASM}
data are plotted against the total intensity. Cyg X-2 data points are averaged
for 3 days and LMC X-3 data points are averaged for 10 days.}
\end{figure}

\section{Discussion}

Apart from the binary period, long term periodic variations are known to be
present in many X-ray binaries. There are four sources in which the presence
of long periods is very well established, (1) Her X-1, a 1.7 day binary with a
35 day period (Giacconi et al. 1973), (2) LMC X-4, a 1.4 day binary with a
30.5 day period (Lang et al. 1981), (3) SMC X-1, a 3.9 day binary with a
long period of 60 days (Wojdowski et al. 1998) and (4) SS 433, a 13.1 day
binary with a long period of 164 days (Margon et al. 1979).
Incidentally, the first three of these sources are also X-ray pulsars. In
SS 433, the periodicities are detected photometrically and spectroscopically
in the optical band only. There is also evidence of periodic component in
several other sources. Among the high mass X-ray binaries, periodicity has been
observed in Cyg X-1 at 300 days (Priedhorsky et al. 1983, also see Kitamoto
et al. (1999) for the {\it GINGA-ASM} observations of a $\sim$150 day period),
4U 1907+09 at 42 days (Priedhorsky  \& Terrell 1984)
and LMC X-3 at 198 (or 99) days (Cowley et al. 1991). Among the low
mass X-ray binaries, Smale \& Lochner (1992) found periodicity in three
sources, Cyg X-2 (78 days), 4U 1820-303 (175 days) and 4U 1916-053 (199 days).
A 106 day periodicity was discovered from an extragalactic point source in
the spiral galaxy M33 (Dubus et al. 1997). The {\it RXTE-ASM} observations of
a large number of sources, for the past three years, detected intensity
variations in many sources (see Levine 1998 for a summary and the light
curves). Some of these sources are of periodic nature and there has also been
discovery of new periodic sources using the {\it RXTE-ASM} data (Sco X-1,
Peele \& White 1996; X2127+119, Corbet, Peele \& Smith 1996). Unstable
long-term periodicity that is attributed to activity of the companion star
or instability of the accretion disk has been observed in Aql X-1 
(Kitamoto et al. 1993). The ratio of the long and orbital periods in
these sources has a wide range between 5 (in 4U 1907+09) and 22,000 in (4U
1820-303). In the two binaries Cyg X-2 and LMC X-3, the reported long term
periodicity is larger by a factor of 8 and 116 than the respective orbital
periods.

The long periods in Her X-1, LMC X-4 and SS 433 are believed to be produced
by the precession of the accretion disks. The mechanisms proposed to cause the
precession of the disks are, (1) forced precession of a tilted disk by the
gravitational field of the companion star (Katz 1973) and (2) precession of a
disk that is slaved to a misaligned companion star (Roberts 1974). But, the
time scale of precession expected in the sources in which binary parameters
are well known is in disagreement with the observed long periods (Priedhorsky
\& Holt 1987). One additional problem is that a large excursion is observed
in the long period of Her X-1 (\"Ogelman 1985) and SS 433 (Margon 1984), and it
may also be present in LMC X-4. This is not explained in the above two models
of disk precession (see Priedhorsky \& Holt 1987, for a detailed discussion).
However, in a realistic case, when various factors like magnetic pressure,
radiation pressure, tidal force, relativistic frame dragging etc. are
considered, it is possible to have significant deviation in the precession
period from its time averaged value. Recent developments in the models,
including tilted
and twisted disks due to coronal winds (Schandl \& Meyer 1994; Schandl 1996)
and warped precessing disks due to radiation pressure (Maloney, Begelman \&
Pringle 1996; Wijers \& Pringle 1998) have a provision for variations in
the precession period. Other models have also been considered to explain
the long periods. These are 1) precession of the compact object (Tr\"umper et
al. 1986), 2) influence of a third body (Fabian et al. 1986) and 3) periodic
modulation of the mass accretion rate (Priedhorsky \& Holt 1987). Among
these models, the first one is not applicable for black hole sources,
and the second one appears to be not true for the two pulsars Her X-1
(Tananbaum et al. 1972) and LMC X-4 (Pietsch et al. 1985). In Her X-1 and LMC
X-4, the presence of a third body required to produce the long period, would
have resulted in additional detectable variations in the pulsation property
(Priedhorsky \& Holt 1987). Periodic and asymmetric mass transfer, which is
induced by the disk's shadowing of the Roche-lobe overflow region, also
contributes to the slaved nature of the accretion disk and is a possible
mechanism. Mass transfer feedback induced by X-ray irradiation may also
generate the observed long term periodicities (Osaki 1985).

No clear understanding of the reason behind the observed
long-term periodicities stands out among all these possibilities.
None of the possibilities mentioned above can explain the long-term
periodicity in all the sources. In the high mass X-ray binaries, the
periodicity is
generally believed to be related to disk precession and in the low mass
X-ray binaries another possibility is some type of disk instability or
modulation in the mass accretion rate related to or induced by the
X-ray radiation (Meyer 1986).

Varying obscuration by the disk (caused by precession), which provides a good
explanation for the long term periodicity, should have the following
observational consequence. With increased absorption in the low intensity
phase, a hardening in the spectrum is expected which is known to be present
in Her X-1 and LMC X-4. The spectral hardening is likely to be more pronounced
between the two lower energy bands. Contrary to this expectation, we find that
in Cyg X-2, HR1 is uncorrelated to the total intensity whereas HR2 is 
anti-correlated, and in LMC X-3, HR1 has a weak positive correlation (Figure
7). Therefore, the spectral variations do not provide very good support to
the disk obscuration scenario.

The present work involving light curves of these two sources with very
large time base suggests that there is no stable periodicity in either of
these systems. There are indications that the oscillatory components seen
in the light curves have varying amplitude and period. Forced or slaved
precession of the accretion disk is unlikely to be the mechanism behind
the quasi periodic intensity variations observed here. Scenarios including
disk precession induced by radiation pressure or tilted and twisted disk
structure induced by wind also require spectral variations different from
the observed pattern. The presence of a third body is also likely to produce
more regular patterns in the intensity variations. Another plausible
explanation for the observed behaviour is instability in the disk or in the
mass accretion rate. But Kuulkers et al. (1996) have pointed out that changes
in the mass accretion rate is unlikely to produce the long term behaviour
observed in Cyg X-2, because changes in the mass accretion rate on a time
scale of less than a day and associated spectral changes are in fact known
to produce the Z pattern (Hasinger \& van der Klis 1989).

However,
if we look only at the {\it RXTE} ASM data of Cyg X-2 (Figure 3), it appears
that there are two unrelated components of intensity variations at periods of
$\sim$40 and $>$60 days, with the first component more stable in period than 
the second one. The results obtained from LMC X-3 (Figures 5 and 6) is somewhat
similar with a relatively stable component at $\sim$100 days and a highly
varying component at $>$130 days. It is possible that in both objects
there are two sources of intensity variation, the first one relatively
stable with a smaller period caused by precession of the accretion disk or by
a third body and the second one unstable in time and having a longer time
scale caused by disk instability or changes in the mass accretion rate.
Coupling between the two components may result in incorrect measurement
of the period of the first component.

\begin{acknowledgements}

We thank an anonymous referee for many suggestions which helped to improve
a previous version of the manuscript.
This research has made use of data obtained through the High Energy
Astrophysics Science Archive Research Center Online Service, provided by the
NASA/Goddard Space Flight Center. We  also thank the {\it RXTE-ASM} and {\it
GINGA-ASM} teams for providing the valuable data. B. Paul was supported
by the Japan Society for the Promotion of Science through a fellowship.

\end{acknowledgements}


\begin{deluxetable}{lccc}
\footnotesize
\tablenum{1}
\tablecaption{Different periods and their significances\label{tbl-1}}
\tablewidth{0pt}
\tablehead{
\colhead{Instrument}&\colhead{Duration}&\colhead{Periods (day)}
&\colhead{Periods in small segments} \nl
\colhead{(energy band)}&\colhead{(days)}&\colhead{(false alarm prob.)}
&\colhead{(false alarm prob.)} \nl
}

\startdata
& & {\bf Cyg X-2} & \nl
\nl
RXTE-ASM & 1078 & 40.4 (1 E-20), 68.8 (7 E-25) & 39.5 (4 E-9), 70.1 (1 E-12) \nl
(1.5--12 keV) & & & 39.6 (9 E-6), 56.3 (7 E-8), 72.4 (3 E-7) \nl
              & & & 41.8 (2 E-7), 56.8 (5 E-7), 73.7 (1 E-10) \nl
GINGA-ASM & 1645 & 53.7 (1 E-4), 61.3 (1 E-7) & \nl
(1--20 keV) & \nl
ARIEL-5 & 1963 & 41.3, 46.2, 53.9\tablenotemark{a}, 77.4 (1 E-16) & 45.7 (3 E-3), 55.3 (1 E-5), 78 (0.9)\tablenotemark{b} \nl
(3--6 keV) & & & 43.2 (0.015), 68.6 (4 E-5), 78 (0.8)\tablenotemark{b} \nl
	   & & & 41 (0.025), 53 (1 E-4), 78 (0.07)\tablenotemark{b} \nl
VELA-5B & 3675 & 33.9 (0.4), 77.4 (0.03) \nl
(3--12 keV) & \nl
\nl
& & {\bf LMC X-3} & \nl
\nl
RXTE-ASM & 1080 & 104 (3 E-38), 169 (2 E-27), 216 (1 E-28) &
104 (3 E-27), 178 (2 E-35) \nl
(1.5--12 keV) & & & 102 (6 E-12), 222 (1 E-11) \nl
GINGA-ASM & 1682 & 105 (1.5 E-3), 214 (1 E-10), 328 (2 E-5) \nl
(1--20 keV) & \nl
HEAO 1 & 506 & 99 (2 E-16), 203 (4 E-12) \nl
(1--13 keV) \nl
ARIEL-5 & 1903 & 96 (8 E-3), 130 (0.03) \nl
(3--6 keV) \nl

\tablenotetext{a}{41.3 (1 E-15), 46.2 (7 E-20), 53.9 (2 E-23)}
\tablenotetext{b}{In the periodograms generated from the segments of
ARIEL 5 light curve, the peaks near 78 days are barely visible, but
in the periodogram of the complete light curve it is highly significant
(see Figure 2).}
\enddata
\end{deluxetable}

\end{document}